\documentclass[iop]{emulateapj}
\usepackage{amsmath}
\usepackage{longtable}
\usepackage{threeparttablex} 

\shorttitle{Tilting Planets}
\shortauthors{Li}

\begin{document}

\title{Tilting Planets during Planet Scattering}
\author{Gongjie Li \altaffilmark{1}}
\affil{$^1$ Center for Relativistic Astrophysics, School of Physics, Georgia Institute of Technology, Atlanta, GA 30332, USA}
\email{gongjie.li@physics.gatech.edu}

\begin{abstract}
Observational constraints on planet spin-axis has recently become possible, and revealed a system that favors a large spin-axis misalignment, a low stellar spin-orbit misalignment and a high eccentricity. To explain the origin of such systems, we propose a mechanism that could tilt the planet spin-axis during planet-planet scattering, which are natural outcomes of in-situ formation and disk migration. Specifically, we show that spin-orbit resonances could occur for a short time period during the scattering processes, and excite the misalignment of the planet spin-axis. This typically leads to planets with large spin-misalignment and a wide range of eccentricities and inclinations. 

\bigskip
\end{abstract}


\section{Introduction}
Tilt of the planetary spin-axis provides important constraints on the formation and the subsequent dynamical evolution of the planetary systems \citep{Lissauer93}. For the Solar System planets, the tilt of the planets have been well explored, and a large number of factors could influence the obliquity (spin-orbit misalignment) of the planets. For instance, the terrestrial planets all encountered chaotic obliquity variations in the past due to secular resonances between planet spin-axis and orbital perturbation \citep{Ward73, Laskar93, Touma93, Li14}, while the spin-axis of Mercury and Venus have been stabilized by tidal dissipation and that of the Earth have been stabilized by the capture of the Moon. For the gas giants, Jupiter's obliquity remains low, while Saturn has a larger obliquity of $26.7^\circ$. Both of them could be affected by resonances with either Uranus or Neptune during planet migration \citep{Ward04, Vokrouhlicky15}. Uranus has a large obliquity above $90^\circ$, and this is likely due to giant impacts as well as spin-orbit resonances in the presence of the circumplanetary disk \citep[e.g.,][]{Slattery92,Morbidelli12, Rogoszinski20}.

Little is known for extrasolar planets, comparing with those residing in the solar system. Theoretical estimates on the obliquity variations of planets have been done for those residing in the habitable region, motivated since obliquity determines the latitudinal distribution of stellar radiation and is important for snowball transition of planets \citep[e.g.,][]{Kane17, Shan18, Saillenfest19, Quarles19, Quarles20}. For instance, it is shown that Kepler 62f and Kepler 186f do not require a massive moon to stabilize their obliquity, different from Earth. In addition, different mechanisms have been proposed to tilt the spin-axes of exoplanets during planet formation, such as via planet-disk interactions \citep{Millholland19, Martin20, Su20}, planet-planet interactions \citep{Ward04, Vokrouhlicky15, Quillen18, MillhollandLaughlin19}, planet collisions \citep{LiLai20} and satellite migration \citep{Saillenfest21}.

Many observational techniques have been proposed to constrain the obliquities of planets~\citep{Barnes03, Gaidos04, Carter10, Schwartz16}. In particular, the first constraint on the orientation of the spin-axis of a planet mass companion has been made for the directly imaged system 2MASS J01225093–2439505 (2M0122), using the projected rotation rates for the companion \citep{Bryan20}. It was found that while the stellar obliquity prefers alignment, the companion obliquity favors misalignment. In addition, the planet mass companion is most likely in a high eccentricity orbit. It was suggested that while secular spin-orbit resonances (due to the low planet spin precession rate), collisions (due to high escape velocity of the massive planet) and Kozai-Lidov oscillations (due to low stellar obliquity) are all unlikely causing the tilt of the planet spin-axis, gravito-turbulent disks provide a promising scenario to tilt the planetary spin-axis. 

To further explore mechanisms that lead to planets with large obliquities, low stellar obliquities and high eccentricities, which are currently out of spin-orbit resonances, we investigated spin-axis variations during planet-planet scattering. Planet-planet scattering naturally leads to orbits with a wide range of eccentricities and inclinaitons. In addition, planet-planet scattering is a common outcome of planet formation, as planets form in a nearly maximally packed configuration in the protoplanetary disk \citep{kokubo02, Goldreich04, IdaLin04}. Once the disk dissipates, mutual planetary perturbations will lead to eccentricity and inclination growth, as well as close encounters between the planets \citep[e.g.,][]{Rasio96, Chambers96, Lin97, Adams03, Ford2008, Chatterjee08, Petrovich14}. It has been found that planetary scattering could lead to high eccentricity planets and planets with large stellar spin-orbit misalignment. More recently, it is responsible for the formation of many of the warm jupiters \citep{Mustill17,Frelilkh19,Anderson20}.

During planet-planet scattering, planets may collide with each other, and the spin-axis of the merger product could be misaligned due to the conservation of angular momentum and linear momentum \citep{LiLai20}. However, the changes in eccentricities and inclinations typically remain low during collisions. Thus, in our paper, we focused on scattering of planets with no collisions. The fraction of systems that avoid collisions increases with increasing planetary semi-major axis and planetary masses.

It is often assumed that the spin-axis have little changes during scattering, since the sizes of the planets are small, and the total torque on the planets during the scattering encounter are negligible to tilt the planet \citep{Lee07}. Nevertheless, here we show that spin-orbit resonances could take place during the scattering events and tilt the planet spin-axis. This could lead to misaligned planets with a wide range of eccentricities and inclinations, as well as lower spin-rate comparing with that due to collisions. This effect is similar to the tilt of the solar system giant planets during planet migration.

The paper is organized as the following: we describe our scattering experiments in section \ref{sec:scat}, where section \ref{sec:setup} presents the set up of the scattering experiments, section \ref{sec:exam} focuses on one illustrate example, and the results of the Monte Carlo simulations of the scattering experiments are summarized in section \ref{sec:scatres}. Then, we explore a larger parameter space and describe the dependence of planetary spin-axes inclination on the properties of the planet system in section \ref{sec:para}. In the end, we summarize the main findings in section \ref{sec:dis}.

\section{Scattering Experiments}
\label{sec:scat}

\subsection{Experiment Setup}
\label{sec:setup}
We run N-rigid-body simulations to study the spin-axis evolution during planet-planet scattering, using the \texttt{GRIT} simulation package \citep{Chen21}. The \texttt{GRIT} simulation package contains a symplectic Lie-group integrator that we developed to simulate systems with gravitationally interacting rigid bodies \citep{Chen21}. Different from the integrator developed by \citet{Touma94}, we do not assume the orbits to be near Keplerian, and thus, this integrator is suitable to study the effects during planet-planet scattering. Tidal dissipation is included in the \texttt{GRIT} simulation package. However, we note that tides make negligible effects in the final distribution of planetary obliquity, and thus we do not include tide for our scattering experiments. General relativistic effects (first-order post-Newtonian) are included in the scattering experiments since planets can be scattered very close to the host star. 

In our default set of simulations (results summarized in section \ref{sec:scatres}), we include three planets in each of the planet system. We set the initial eccentricities of the planets to range between $0$ and $0.1$ distributed uniformly and the initial inclination to range between $0^\circ$ and $2^\circ$ uniformly. Higher initial eccentricities allow instabilities to occur faster. We vary the initial orbital parameters (e.g., changing the initial eccentricities to range between $0$ and $0.05$ in section \ref{sec:para}). We set the planets to be separated by $3.5R_{H, mutual}$ from each other, where the mutual Hill radius is expressed as the following:
\begin{align}
    R_{H, mutual} =\Big(\frac{m_1+m_2}{3M_{star}}\Big)^{1/3} \frac{a_1+a_2}{2} ,
\end{align}
where $m_1$ \& $m_1$ and $a_1$ \& $a_2$ are the masses and semi-major axes of the two planets separately, and we placed the innermost planet at $1$AU. The argument of pericenter, longitude of ascending node and mean anomaly of the planets are uniformly distributed between $0$ and $2\pi$.

For each of the planets, we set their masses to range between $1 m_{jup}$ and $2 m_{jup}$ distributed uniformly. The mass range is set arbitrarily, and we discuss the effects with different planet masses in section \ref{sec:para}. We set the radius of the planets to be one jupiter radius, and the spin rate of the planets to be $30\%$ of the break up spin rate. The oblateness of the planets are obtained using Darwin-Radau relation \citep{MurrayDermott00}:
\begin{align}
    \mathbb{C} \equiv \frac{C}{m_pR_p^2} = \frac{2}{3}\Big[1-\frac{2}{5}
    \Big(\frac{5}{2}\frac{q}{f}-1\Big)^{1/2}\Big] ,
\end{align}
where $C$ is the planet's moment of inertia around the rotational axis, $m_p$ is the mass of the planet, $R_p$ is the radius of the planet, $q$ represents the relative importance of the centripetal acceleration and $f$ represents oblateness. We assume the moment of inertia coeffient $\mathbb {C} = 0.26$ for gas giants and obtain $f=0.067$. This is similar to the oblatess of Jupiter ($f_{jup} = 0.06487$), and smaller than that of Saturn ($f_{sat} = 0.09796$). We assume the love number of the gas giants to be $k_2 = 0.5$, following the $n = 1$ polytrope density distribution. We assume the stellar $J_2$ moment to be $2\times10^{-7}$, similar to the Sun, though we note that the stellar $J_2$ moment has little effects on the final distribution of the planetary spin-axes. 

We run the simulations for 1Myr ($10^6$ times the innermost planet's period). This is much longer than the instability time of the planets with a separation of $3.5 R_{H, mutual}$ ($\sim 300$yr) \citep{Chatterjee08}. The time step is set to be $10^{-4}$yr, less than $5\%$ of the spin period of the planets.

\subsection{Illustrative Example}
\label{sec:exam}

\begin{figure}[h]
\center
    \includegraphics[width=0.45\textwidth]{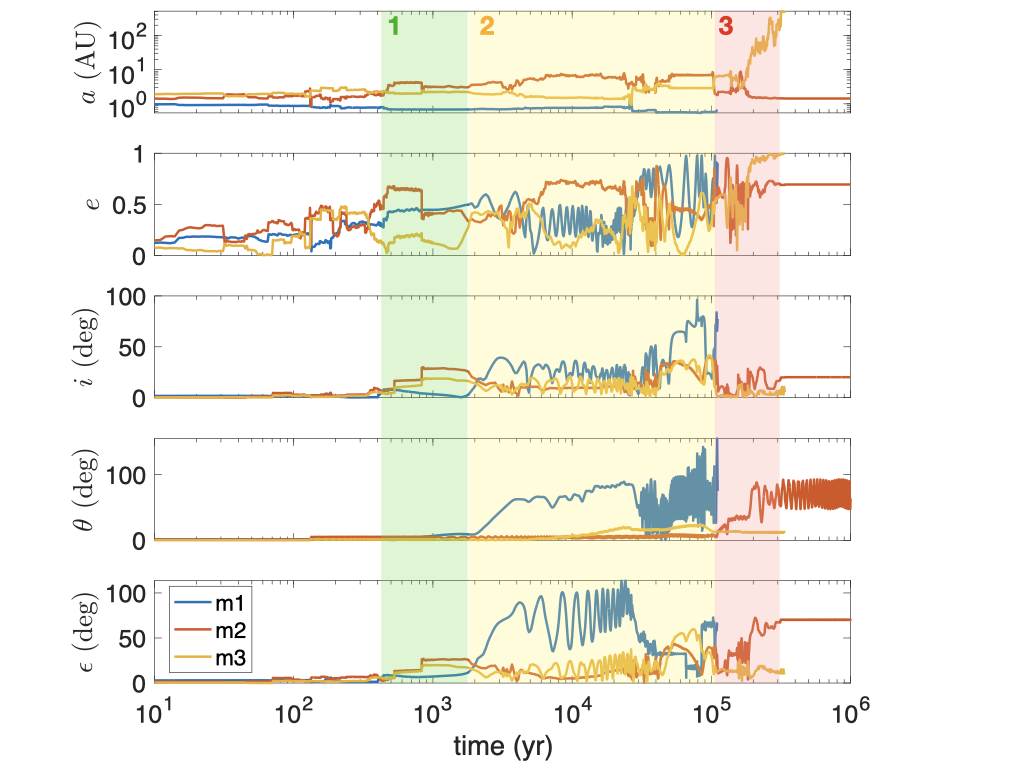}
    \caption{Illustrative example of a scattering experiment. $\theta$ and $\epsilon$ represent the tilt of the spin-axis relative to the reference plane and the spin-orbit misalignment (obliquity) separately. Both planet $m_1$ and $m_2$'s spin axes are largely tilted during the scattering process, and planet $m_2$'s inclination remains low $\sim 20^\circ$ after the encounter.  \label{fig:illu}}
\end{figure}

To illustrate how the planetary spin-axes are tilted during planet scattering, we include here an illustrative example using one of the Monte Carlo draws. The mass of the planets are $1.1643m_{jup}$, $1.0090m_{jup}$ and $1.6349m_{jup}$ separately, and the initial eccentricities and inclinations are $0.052255$, $0.098151$ and $0.091897$, as well as $1.4622^\circ$, $0.9179^\circ$ and $0.3547^\circ$ separately. Only one of the planet ($m_2$) survived after the close encounters. The orbital evolution of the three planets are shown in Figure \ref{fig:illu}.

As shown in Figure \ref{fig:illu}, eccentricities of all the planets are excited to large values during the scattering, while inclinations of $m_2$ and $m_3$ remain low. Both the spin-axes of $m_1$ and $m_2$ are tilted. In particular $m_1$ has large and fast spin inclination ($\theta$) variations before it is ejected (phase 2, yellow region). This is because inclination of $m_1$ is excited above $50^\circ$, and spin-axis precession around the orbit drives large amplitude spin-axis variations. In addition, the semi-major axis of $m_1$ becomes small before the ejection, and this leads to fast spin-precession rate and thus fast oscillations in $\theta$. 

On the other hand, $m_2$'s obliquity ($\epsilon$) is slightly increased during phase 1 (green region) due to orbital inclination increase. Then, its spin-inclination ($\theta$) is largely tilted during the second close encounter of the planets (phase 3, red region around $\sim 0.1$Myr to $\sim 0.2$Myr), while the orbital inclination remains low. After the ejection of $m_3$, the obliquity of $m_2$ remains constant, and the spin-axis inclination variations are caused by spin-axis precession around the orbital normal direction due to spin-orbit coupling. 

\begin{figure}[h]
\center
    \includegraphics[width=0.45\textwidth]{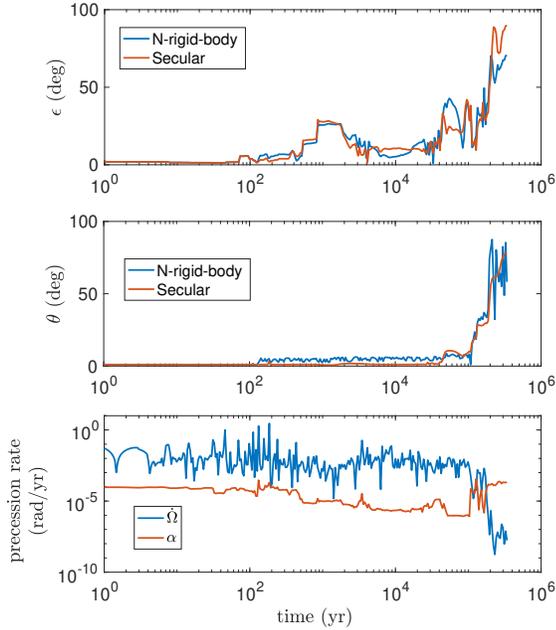}
    \caption{Obliquity and spin evolution (top two panels) compared with the spin and orbital precession rate (bottom panel). In the top two panels, the blue lines represent the results of the N-rigid-body simulation, and the red lines represent the secular results. The secular results largely agrees with the N-rigid-body simulations, and disagreement is expected since secular approximation is not valid during planetary close-encounter when the orbital elements have rapid changes. The bottom panels shows the precession rates. It illustrates that when the spin-precession rate coincides with that of the orbital precession, the spin-axis of the planet is tilted. \label{fig:127spin}}
\end{figure}

How is the spin-axis of $m_2$ tilted? To investigate the mechanism that tilt the spin-axis of the surviving planet $m_2$, we plot the obliquity and spin evolution using both the N-rigid-body simulation and the secular approach. We note that it is not appropriate to use secular method to estimate the evolution of the system during planet-planet scattering, where the semi-major axes change drastically. However, we use it as a probe to illustrate the effects of the non-adiabatic resonance between the spin and orbital precessions that enhanced spin-inclinations. This is analogous to the secular spin-orbit resonance encounters during planet migration \citep[e.g.,][]{Vokrouhlicky15, MillhollandLaughlin19}, while the semi-major axes changes can be more drastic during scattering. We plot the precession rate of the spin-axis and that of the orbital longitude of ascending node in Figure \ref{fig:127spin}. The secular results are obtained following the Hamiltonian listed below \citep[e.g.,][]{Laskar93}:
\begin{align}
    H(\xi, \psi, t) = \frac{1}{\alpha}\xi^2 + \sqrt{1-\xi^2}(A(t) \sin{\psi}+B(t) \cos{\psi}) , 
\end{align}

where $\xi = \cos{\epsilon}$ and $\psi$ is the spin-longitude projected in the orbital plane, and $\alpha$ is the spin-precession coefficient \citep[e.g.,][]{Ward04}:
\begin{align}
    \alpha &= \frac{k_2}{2\mathbb{C}}n\Big[\frac{\omega R_p^3}{\omega_b a^3 (1-e^2)^3}\Big]^{1/2} , \\
     & =5.9\times10^{-4}{\rm yr^{-1}} \nonumber \\
    &~~ \times \Big(\frac{k_2/(2\mathbb{C})}{0.5/0.52}\Big)\Big(\frac{n}{2\pi yr^{-1}}\Big)\Big[\frac{\omega/\omega_b}{0.3}\frac{(R_p/R_{jup})^3}{(a/1AU)^3 (1-e^2)^3}\Big]^{1/2} \nonumber
\end{align}
where $n$ is the orbital frequency $\omega$ and $\omega_b$ are the spin velocity and the breakup velocity of the planet. In addition, $A(t)$ and $B(t)$ reflect the orbital variations and are expressed as the following:
\begin{align}
    A(t) = 2(\dot{q} + p(q\dot{p}-p\dot{q}))/\sqrt{1-p^2-q^2} \\
    B(t) = 2(\dot{p} - q(q\dot{p}-p\dot{q}))/\sqrt{1-p^2-q^2} ,
\end{align}
where $p = \sin{i}/2\sin{\Omega}$ and $q = \sin{i}/2 \cos{\Omega}$. We obtain the orbital evolution using the results of the N-rigid-body simulations and integrate the spin-evolution using the secular approach before the ejection of $m_3$.

For the precession rates, the orbital precession time is directly obtained taking the time derivative of the longitude of ascending node ($\Omega$), and we use the spin-axis precession coefficient $\alpha$ as a proxy for the spin-axis precession rate.

As shown in Figure \ref{fig:127spin}, obliquity of the planet is increased during the first close encounter between the planets around $\sim 1$kyr. The inclination of $m_2$ is excited, while the spin-axis inclination ($\theta$) is not affected significantly during the close encounter. Thus, obliquity is increased due to changes in the orbital inclination. Next, at $\sim 0.1$Myr, after the close-encounter between $m_2$ and $m_3$, the semi-major axis of $m_2$ is reduced, which leads to faster spin-precession rate. Meanwhile, the orbital precession rate decreases as the planet companion $m_3$ scatters with $m_2$ and migrates outward. This causes sweeping of spin-orbit resonance, which drives obliquity excitation and the tilt of the spin-axis while the inclination of the orbit remains low ($\sim 20^\circ$).

\subsection{Scattering Results}
\label{sec:scatres}

We run $500$ scattering experiments to study the distribution in the tilt of the planetary spin-axes during planet-planet scattering. We sample the orbital parameters and masses of the planets as described in section \ref{sec:setup}. $\sim 60\%$ of the scatterings avoid planet-planet collisions. We focus on systems that have no collisions, and the results are summarized in this section.

\begin{figure}[h]
\center
    \includegraphics[width=0.45\textwidth]{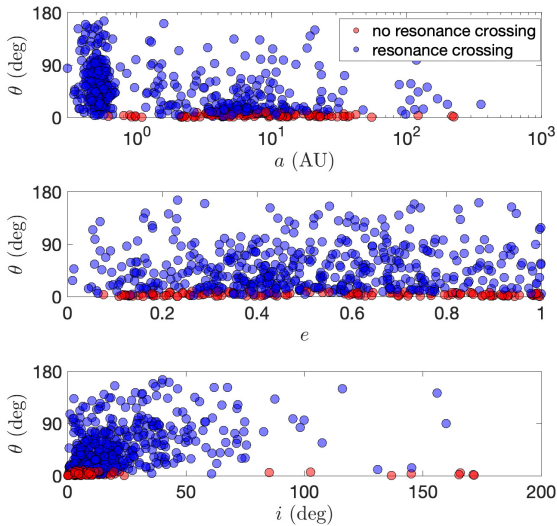}
    \caption{Results of scattering experiments. Tilt of planet spin-axis as a function of planet semi-major axis (top panel), eccentricity (middle panel) and inclination (bottom panel). circles mark the runs that did not encounter spin-orbit resonance crossing, and blue circles mark those encountered the resonance crossing. Eccentricity and inclination can be excited to large values during planet scattering, and there is a wide spread in eccentricity and inclination for large tilt of planetary spin-axis. It is more easily to tilt planet spin-axis with larger planet inclination, due to precession of the spin-axis around the orbital normal. \label{fig:scatterF3}}
\end{figure}

Figure \ref{fig:scatterF3} shows the tilt of the spin-axis with respect to semi-major axes, eccentricities and inclinations of the planets. The dearth of planets around $\sim 1$AU is due to planet-planet scattering as the innermost planets start at $1$AU. There is a wide spread in semi-major axes, eccentricities and inclinations for planets with large spin-axis inclination ($\theta$). The dependence of of the spin-axis inclination on the final eccentricity of the planets is weak, but it is more likely to tilt the spin-axis of the planets with larger final inclinations. This is because spin-axis precession around the orbit leads to larger spin-axis inclination when the orbit is more inclined relative to the reference plane. In addition, the obliquity oscillation amplitude is larger with higher inclination under spin-orbit resonances \citep[e.g.,][]{Shan18}. 

To investigate the role of spin-orbit resonances, we mark the runs that encountered spin-orbit resonance crossing (with matching spin-precession frequency and orbital precession frequency) in blue. Spin-orbit resonances are prevalent during scattering. Specifically, $80.2\%$ of the systems encountered spin-orbit resonances, and all systems with spin-inclination excited above $10^\circ$ encountered spin-orbit resonance crossing. Systems with lower semi-major axes have a larger probability to encounter spin-orbit resonances ($97.8\%$ for $a<1$AU) comparing with their farther companions ($66.0\%$ for $a>1$AU). Some systems have spin-inclination enhanced due to both orbit precession and spin-orbit resonances. Overall, it is more likely to tilt the planetary spin-axis with low semi-major axis due to faster spin-precession rates that are both more likely to commensurate with that of orbital precession, and to tilt the spin-axis around orbital normal more efficiently. 

Precession timescales are longer for farther companions (e.g., $\sim 10$Myr at $10$AU for Jupiter-like planet), and this will enhance planetary spin-inclination over long timescales when planetary spin-axes precess around inclined orbits. We note that we only focus on the outcomes shortly ($1$Myr) after planet-planet scattering, and the detailed long term dynamics will be discussed in a follow up study.

\section{Parameter Studies}
\label{sec:para}

\begin{figure}[h]
\center
    \includegraphics[width=0.5\textwidth]{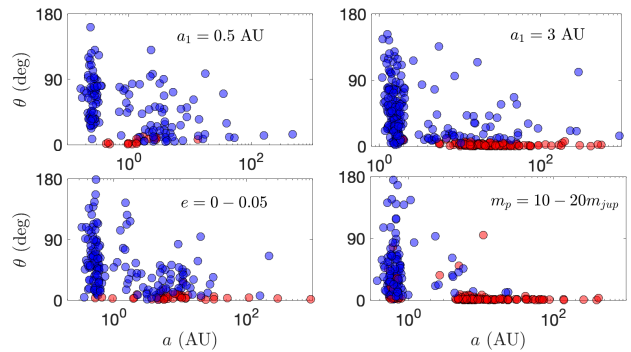}
    \caption{Tilt of planet spin-axis as a function of planet semi-major axis for different sets of scattering experiments. The set up of the simulations are the same as that described in section \ref{sec:setup}, except that in the {\it upper left panel}: innermost planet starts at 0.5 AU; {\it upper right panel}: innermost planet starts at 3AU; {\it lower left panel}: initial eccentricities range between 0 and 0.05; and {\it lower right panel}: planet masses range between $10-20m_{jup}$. Similar to Figure \ref{fig:scatterF3}, red circles mark the runs that did not encounter spin-orbit resonance crossing, and blue circles mark those encountered the resonance crossing. It is more easily to tilt planet spin-axes for close-in planets for all the scattering experiments, and the planetary spin-axes can be tilted to larger values for planets with higher $v_{orb}$ to $v_{esc}$ ratio. \label{fig:par}}
\end{figure}

We consider different scattering experiments varying the distribution of the initial eccentricity, semi-major axes, as well as the mass of the planets in this section. For each set of simulations, we run 200 experiments. The distributions of the spin-axes inclinations as a function of planet final semi-major axes are shown in Figure \ref{fig:par}.

The upper panels of Figure \ref{fig:par} include different initial semi-major axes of the innermost planets comparing with the default set of simulations in section \ref{sec:scatres}. As the planets reside farther from the host star, the orbital velocity decays, and planets can avoid collisions more easily (with lower ratio of the orbital velocity to the escape velocity, see e.g., \citealt{Petrovich14}). However, the spin-axis inclination of the planets can be more easily excited, when the planets start closer to the host stars due to faster spin precession rates. Thus, less planets have their spin-axes tilted during scattering when starting farther from the host star, even though more of them avoid collisions. 

\begin{figure}[h]
\center
    \includegraphics[width=0.45\textwidth]{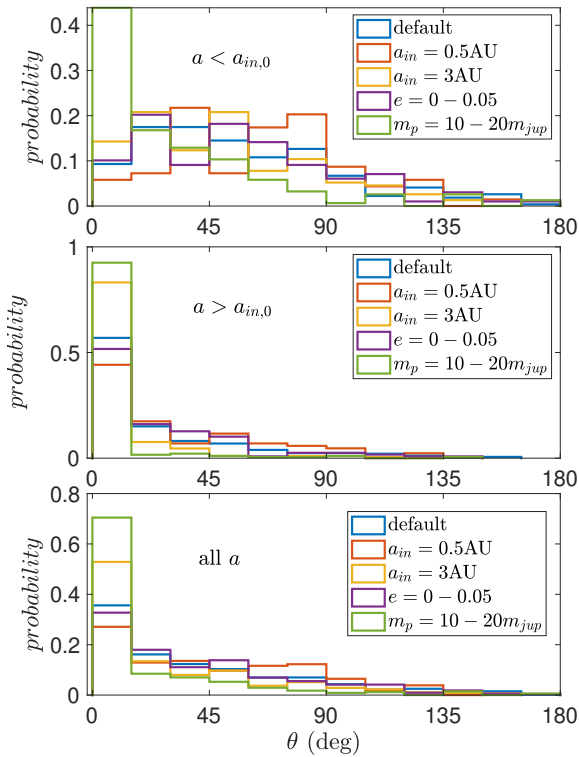}
    \caption{Histogram of the spin-axes inclination for planets with final semi-major axes interior (upper panel) and exterior (lower panel) to the initial semi-major axes of the innermost planets. Planets with higher $v_{orb}/v_{esc}$ ratio residing closer to the host star can more easily tilt their spin-axes. \label{fig:hist}}
\end{figure}

The lower left panel shows the case when the planets start less eccentric ($e$ ranging between $0$ and $0.05$). It takes more time for the eccentricity to grow larger due to planet interactions, and orbital instability occurs at a later time. Nevertheless, spin-axis inclination distribution and its dependence on semi-major axis are similar comparing to the default set of simulations in section \ref{fig:scatterF3}. Note that the distribution of the spin-axis inclination is shown more clearly in Figure \ref{fig:hist}.

When the masses of the planets get larger, the escape velocities increase, and planets experience faster ejection. Thus, as shown in the lower right panel of Figure \ref{fig:par} with planet masses ranging between $10m_{jup}$ and $20m_{jup}$, less number of planets have their spin-axes tilted to large values. 

Similar to Figure \ref{fig:scatterF3}, we mark the runs that encountered spin-orbit resonance crossing in blue. It is more likely to encounter spin-orbit resonances when the planet starts closer to the host star due to the faster precession rates ($89.0\%$ starting at $0.3$AU versus $59.6\%$ starting at $5$AU). Larger planetary masses ($10m_{jup}$ and $20m_{jup}$) are less likely to encounter spin-orbit resonances due to faster ejection processes ($37.9\%$), and lower initial eccentricity ($e=0-0.05$) do not change the probability to encounter the spin-orbit resonances significantly ($79.2\%$). As expected, spin-inclinations are more likely to be enhanced for systems that encounter spin-orbit resonance crossing.

For all the sets of simulations, planets with lower final semi-major axes can have their spin-axes tilted more easily immediately (1 Myr) after the scattering. This is both due to the higher probability to cross spin-orbit resonances closer to the host star and due to the stronger non-adiabatic limit farther from the host star that cannot efficiently excite spin-inclination. Thus, in Figure \ref{fig:hist} we divide the sample to two groups: those with final semi-major axes lower than the initial semi-major axes of the innermost planets (upper panel), and those that reside farther (middle panel). 

For close-in planets, most of them could have large spin-axis inclination, except those with large masses (or low $v_{orb}/v_{esc}$ ratio). Those that initially reside closer to the host star can have spin-axes excited to larger values. The differences in the initial eccentricities do not make significant influence in the distribution of the final spin-axis inclinations. 

For planets at large separation from their host star, a majority of them still have low spin-axis inclination even though some planets could have their spin-axes inclination excited. Comparing between different initial conditions of the scatter experiments, the difference in the initial eccentricity also makes little effects in the final distribution of the spin-axes inclination. When the planetary masses are higher or start farther from the host star (lower $v_{orb}/v_{esc}$ ratio), it is more difficult to excite the planetary spin-axes due to faster planet ejection processes.

Combining the two groups together (bottom panel), $\sim 50\%$ of the scattering experiments produce high spin-axis inclinations over $50^\circ$ when the planets start at $0.5$ AU, and $\sim 20-30\%$ produces high spin-axis inclinations starting at $1-3$AU. Only around $\sim 10\%$ reach high spin-axis inclination for massive planets ($10-20m_{jup}$). 

After the completion of this paper, we became aware of a complementary study that arrived at the same results independently and simultaneously based on two-planet scattering \citep{Hong21}. Our results about the dependence of planet spin-axis inclination on semi-major axes agree with each other. In addition, our results show that three-planet scatterings could lead to a larger number of planets with higher obliquity due to longer instability time and higher inclination excitation.

\section{Conclusion and Discussions}
\label{sec:dis}

In this letter, we discuss planetary spin-axes variations during planet scattering. We find that temporary spin-orbit resonances could lead to large tilt of planetary spin-axes for planets with a wide range of eccentricities and inclinations. It is more likely to tilt planetary spin-axes for planets with lower masses and initially reside closer to the host star before scattering occurs (higher $v_{orb}/v_{esc}$ ratio), which could allow slower ejection processes and longer time to tilt the planets. In addition, it is more likely to tilt the spin-axes of the planets that reside closer to the host stars after planet-planet scattering. 

We note that the spin-axis inclination of the massive planet in system 2M0122 is unlikely due to spin-orbit resonances during planet scattering. Orbiting around a $0.4$M$_\odot$ star, the escape velocity of the $12-27 m_{jup}$ planet greatly exceeds the orbital velocity. The ejection process is fast and could quench obliquity excitation due to secular resonances. Thus, it is challenging to produce planets with large obliquities in particularly at distances around $\sim 50$ AU. It is likely that the tilt of the planet is due to turbulent accretion or the tilt of circumplanetary disk \citep{Bryan20, Martin20}. In addition, as a caveat, recent SPH simulations show that collision rate could be further increased, and this will lower the probability to tilt planetary spin-axes during planet-planet scattering \citep{Li21}.

Many mechanisms have been proposed to explain the origin of tilted planetary spin-axes. For instance, planet-disk interactions, turbulent accretion to planets as well as collisions during planet-planet scattering can also lead to misaligned planetary spin-axes. To distinguish from the other scenarios, planet-planet scattering (with no collisions) could lead to planets with a wide range of eccentricities and inclinations. In addition, planetary spin-rate is nearly unchanged, different from those produced by planet collisions. Obliquity presents an exciting and unique window into formation history, and more systems with measured obliquities will enable statistical studies to disentangle different planet formation mechanisms in the future.

\section*{Acknowledgments}

The author thanks Fred Adams, Smadar Naoz, Konstantin Batygin, Zeeve Rogoszinski and Dong Lai for reading the manuscript and providing helpful comments. The author also thanks the referee for helpful comments, which greatly improved the quality of the letter. GL is grateful for the partial support by NASA 80NSSC20K0641 and 80NSSC20K0522.




\bibliographystyle{hapj}
\bibliography{ref.bib}

\end{document}